\documentclass{ws-p8-50x6-00}
\begin{document}
\title{Future Low $x$ Physics and Facilities}
\author{M. Klein}
\address{DESY Zeuthen \\E-mail: klein@ifh.de}
\maketitle \abstracts{A brief overview is given on the physics and
  future lepton-nucleon collider facilities to explore the domain of
  high parton densities at low Bjorken $x$, with HERA, the
  electron-ion collider (EIC) and with  THERA - in $ep$,
  polarised $\overrightarrow{e} \overrightarrow{N}$ and $eA$ mode.}
\section{HERA-2}
The only existing lepton-nucleon collider which accesses the region of
low Bjorken $x$ is HERA. During 2001 the detectors were upgraded and
the machine was modified to significantly increase the luminosity.
Spin rotators were installed enabling to run the collider experiments
H1 and ZEUS with longitudinally polarised positrons and electrons.  It
is now expected to collect over the coming five years an integrated
luminosity of 1~fb$^{-1}$, which represents about ten times the
luminosity recorded in HERA's first running period. This luminosity
upgrade primarily is dedicated to the exploration of the low cross
section region at high momentum transfers, $Q^2 \simeq M_Z^2$.  Yet, a
wealth of precise low $x$ data can still be expected from HERA which
are necessary to further develop low $x$ theory. These comprise for
example: i) a precision measurement of $\alpha_s$ from scaling
violations of $F_2$, ii) the accurate measurement of the $x$
dependence of the longitudinal structure function $F_L$ from a series
of runs with lowered proton beam energy, iii) first measurements of
$F_2^b$ and precision data on $F_2^c$ in an extended $x$ range using
new Silicon detectors, iv) data on the diffractive structure functions
$F_2^{D3}$ and $F_2^{D4}$ with a new, very forward proton
spectrometer, v) accurate data on vector meson production, on final
states and on charm in diffraction.  An exciting decade of low $x$,
high precision $ep$ physics is ahead.  Key questions in low $x$
physics, however, will remain unanswered even after completion of the
HERA-2 run because it is restricted to electron-proton scattering at
energy $s =4E_eE_p < 10^5$~GeV$^2$, and because the low $Q^2$ acceptance
range, around 1~GeV$^2$, in H1 and ZEUS is now obstructed by
focusing magnets.
\section{Low $x$ Physics Beyond HERA-2}
Future low $x$ physics, which can only be sketched here, regards the
unknown behaviour of the neutron, photon and nuclear structure, of the
nucleon spin composition in this domain and it needs to clarify
experimentally the question of saturation. Super-high-energy neutrino
scattering~\cite{neuastro} and the LHC data cannot be reliably
understood without extending the $x$ range considerably beyond the
region accessed in $ep$ scattering at HERA.

After completion of the currently approved HERA programme there remain
precision measurements to be performed in the low $x$ region at HERA.
These regard in particular the measurement of jets very close to the
proton beam direction, to understand the emission of gluons, and the
low $Q^2 ~\sim 1$~GeV$^2$ region, in particular the longitudinal
structure function $F_L$ and the diffractive structure function
$F_2^{D,4}$ \cite{acald}.  Such a programme needs the forward and
backward regions very close to the beam pipe at HERA to be reopened
and upgraded. Efficient proton tagging and a small beam divergence
are required as for electron-deuteron scattering.

Deuterons are a source of quasi-free neutrons, apart
from nuclear shadowing effects at low $x < 0.1$. 
These are related to the diffractive
parton densities which provide enough constraint to control shadowing
to better than 1-2\% of the cross section~\cite{fsdiff}. The
asymmetry of sea quarks, $\overline{u} - \overline{ d}$, can thus be
accurately measured from the difference $F_2^p - F_2^n$, which is free
of Pomeron exchange.  Furthermore, $ed$ data are essential in the
unfolding of parton distributions, such as $s-c$ or the $d/u$ ratio,
in determining charged current structure functions and in precision
tests of the $Q^2$ evolution in perturbative QCD. Electron-deuteron
scattering appears to be the natural next step beyond $ep$ scattering.

Investigation of the nucleon spin structure with $\overrightarrow{e}
\overrightarrow{N}$ colliders will explore spin phenomena at high
$Q^2$ and at low $x$ for the first time. This opens a completely new
field for testing QCD,\hspace*{-2mm}~\cite{adr,terry} making use of the complete
final state reconstruction capabilities. For example, dijet
signatures can be used to measure the
spin distribution $\Delta G$.  The behaviour of the spin
structure function $g_1(x,Q^2)$ at low $x$ is unknown, but expected to
change even more dramatically than $F_2$. The $Q^2$ dependence of
$g_1$ determines the gluon spin distribution $\Delta G$. The integral
of $\Delta G$ enters the proton spin decomposition, $1/2 = J_g + 1/2
\cdot \Delta \Sigma + L_q$, together with the quark orbital angular
momentum and the quark spin distributions, $\Delta \Sigma = \sum_i
\Delta q_i$, which are similarly undetermined at low~$x$.
Here $J_g = \int{\Delta G dx} + L_g$.  Semi-inclusive measurements
explore the flavour structure of spin and DVCS is sensitive to off
diagonal parton distributions. Photoproduction gives insight into the
polarised gluon structure of the photon.  The kinematic region of
polarised $eN$ scattering is a vast terra incognita beyond the fixed
target experiment domain.  The first $ \overrightarrow{e}
\overrightarrow{N}$ collider experiment, extending also to high $Q^2$,
may change radically our view on nucleon spin structure since it 
accesses the sea and the gluon.

\begin{figure}[t]
\begin{picture}(200,142)
\put(-5,-10){
\epsfig{figure=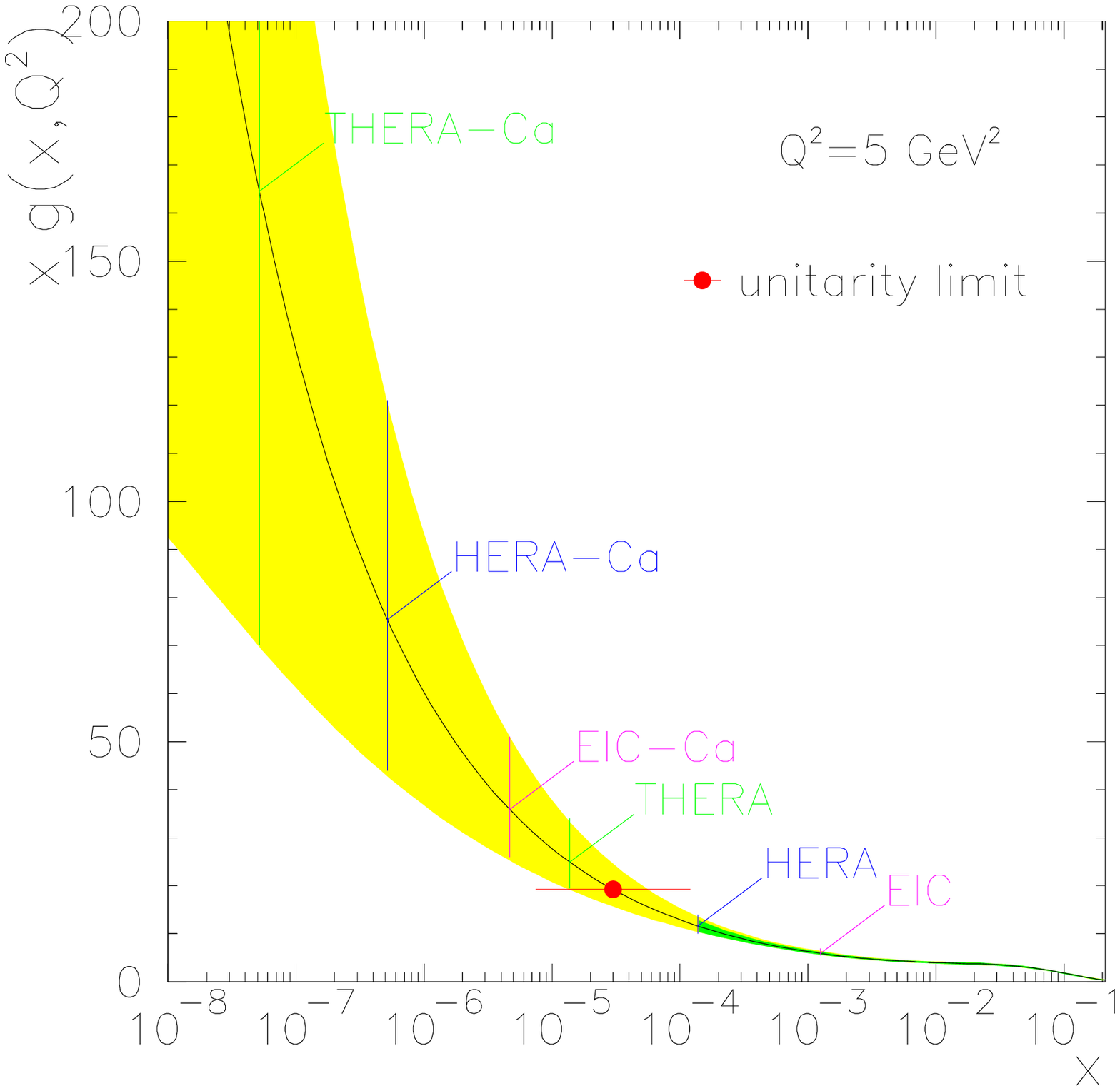,height=5.3cm}}
\put(165,-10){
\epsfig{figure=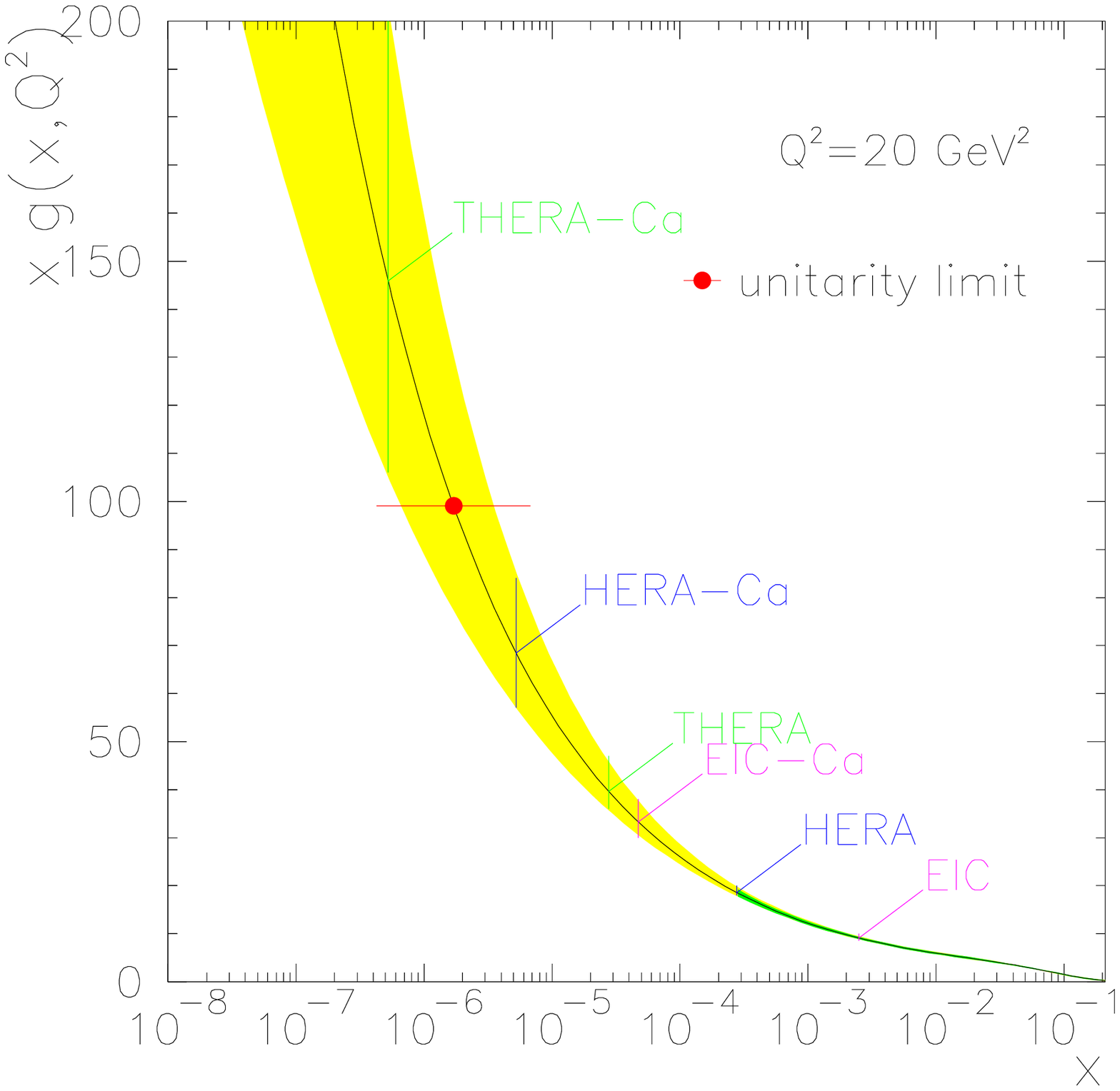,height=5.3cm}}
\end{picture}
\caption{Extrapolation of the gluon distribution 
towards much lower $x$. This $qualitatively$ illustrates the range
accessible with different colliders and $eA$ scattering by comparing the
possible further rise of $xg$ with unitarity limits.
The behaviour of $xg$ at low $x$ may yet differ very much
from this bold extrapolation for various reasons, see text. \label{eAg}}
\end{figure}

The rise of the proton structure function $F_2$ towards low $x$ in the
DIS region is due to a high sea quark density related to a gluon
distribution $xg \propto \partial F_2 / \partial \ln Q^2$ which also
rises towards low $x$ as $x$.\hspace*{-2mm}~$^{-\lambda}$ 
\hspace*{1mm}Since $xg$ is roughly
expected to be enhanced $\propto A^{1/3}$ in nuclei of atomic number
$A$, in electron-nucleus ($eA$) scattering an equivalent Bjorken $x =
x_N /(A^{1/3})^{1/\lambda}$ is accessed: nuclei can thus be utilized
to $preview$ a kinematic region which in $ep$ scattering requires a
considerable increase of energy $s$. This is illustrated by
extrapolating $xg$ and its experimental uncertainty, using the NLO QCD
analysis by the H1 Collaboration~\cite{h1qcd}, much beyond its range
of validity towards extremely low $x$ in Fig.\ref{eAg}. At a certain
value of $x$, the rise of the gluon distribution is expected to
contradict unitarity limits~\cite{thera} which, within an
uncertainty of about a factor of two, limit $xg$ by $Q^2/\alpha_s(Q^2)$.
Depending on higher order corrections, contributions $\propto \ln1/x$,
shadowing effects, the chosen parameterisation of $xg$ at minimum
$Q^2$, the heavy flavour treatment and for different $A$ than Ca
($A=40$), this picture may be drawn differently. Yet, Fig.\ref{eAg}
illustrates the expected density amplification effect of nuclei. Notice in
particular the great extension of kinematic range with
THERA,\hspace*{-2mm}~\cite{thera} which is the $eN$ collider using HERA~($N$) and TESLA
($e^{\pm}$). With nuclei, THERA directly accesses the range of
super-high-energy neutrino astrophysics with $x$ as small as $10^{-8}$.
Diffraction in nuclei may constitute up to 50\% of the inclusive cross
section, and since $xg$ enters squared into the diffractive cross
section, non-linear damping effects may possibly be seen in diffraction
first.
\section{Lepton-Nucleon Collider Prospects}
HERA is scheduled to run with upgraded luminosity of $7
\cdot 10^{31}$cm$^{-2}$s$^{-1}$ until the end of 2006.\hspace*{-1.5mm}~\cite{durhera} 
\hspace*{1mm}The ultimate
value is estimated to be $1.3 \cdot 10^{32}$cm$^{-2}$s$^{-1}$ which
corresponds to an annual luminosity of about 500~pb$^{-1}$.  With
such high luminosity the measurement of small asymmetries at low $x$
in polarised $\overrightarrow{e} \overrightarrow{p}$ scattering is
statistically feasible.  Polarised electron-proton interactions require a high
proton polarisation transferred through the accelerator chain,
polarimeters and Siberian snakes for spin rotation. Because of
the small anomalous magnetic deuteron moment, polarised
$\overrightarrow{e} \overrightarrow{d}$ collisions may possibly be
realised more easily than $\overrightarrow{e} \overrightarrow{p}$.  Ions
can be accelerated at HERA with some modifications regarding the
source, electron cooling in PETRA for heavier nuclei to
counter intrabeam scattering (IBS) effects, longer bunch trains and
ramp cycles.  The luminosity is then estimated to scale as $L_A \simeq
L_p /A$. For deuterons the IBS time exceeds 2h and a luminosity of
$3.5 \cdot 10^{31}$cm$^{-2}$s$^{-1}$ can be achieved without cooling.
Thus a high luminosity $ed$ run can be done right after completion of
HERA-2 without any major modification beyond the source.

The EIC~\cite{ilan} is a project to be proposed in about 2005 as an
intense polarised electron-ion collider using electron-beam-cooled
ions in RHIC (of up to $E_p =250~$GeV proton and $E_{Au} =100~$GeV/A
gold energy) colliding with electrons from a ring or linear
accelerator of 10~GeV maximum energy. Thus the EIC has 10 times less
energy $\sqrt{s}$ than HERA. With high luminosity it yet extends the range of
current polarised experiments and with heavy nuclei it accesses the
region of high parton density.  The ring-ring accelerator has an
estimated luminosity of $25~(0.7) \cdot 10^{31}$cm$^{-2}$s$^{-1}$ for
p (Au) which is comparable to HERA after run 2. With an energy
recovery linac one expects to obtain still higher luminosity of $100
~(1.0) \cdot 10^{31}$cm$^{-2}$s$^{-1}$ for p (Au). The linac, while
accelerating $e^-$ only, has the further advantages of providing high
polarisation and avoiding synchrotron radiation background.

\begin{figure}[t]
\begin{picture}(200,240)
\put(-35,250){
\psfig{figure=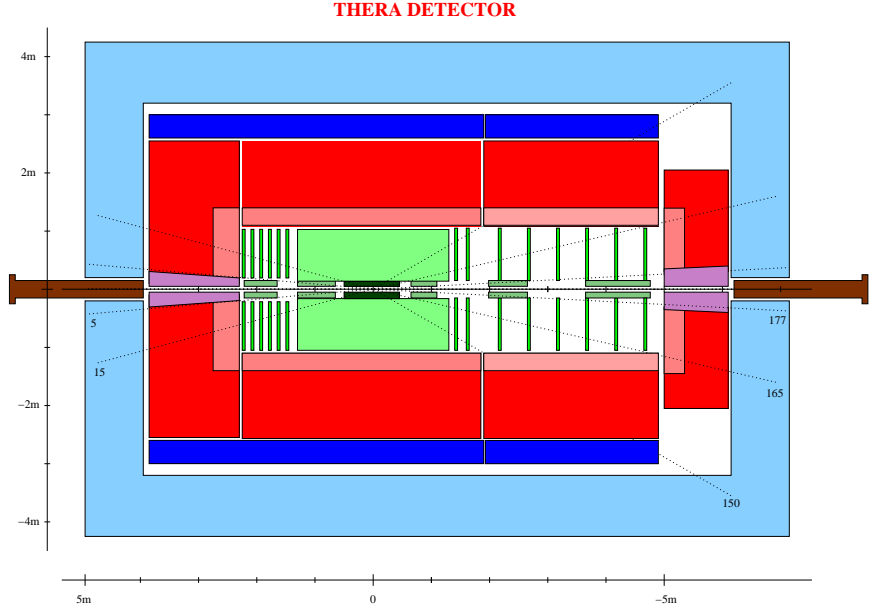,height=14cm,angle=-90.}}
\end{picture}
\caption{Feasibility design of a detector to operate at the $ep$ collider
  THERA in the TeV range of energy. The detector is equipped with
  silicon tracking and plug calorimeters in $p$ and $e$ beam direction,
  in order to measure forward going jets
  and backward scattered low $x$ electrons very closely to the beam axis.  
  It can be
  supplemented with tagging detectors for forward $p,~n$ and backward
  $e,~\gamma$.  The radial dimensions are determined by the proton
  beam energy. Thus a THERA detector resembles H1 or ZEUS with
  tracking, calorimetry, solenoid magnetic field and muon detection.
  \label{tdet}}
\end{figure}

THERA~\cite{thera} collides TESLA leptons of $250-800$~GeV off HERA
nucleons with luminosities of $0.4-2.5 \cdot 10^{31}$cm$^{-2}$s$^{-1}$
corresponding to annual integrated values between 40 and 250~pb$^{-1}$
since in a linear electron machine the current does not degrade with
time. The TESLA cavities are of standing wave type which allows both
arms to be used for acceleration.  The rather symmetric $e/p$ energy
ratio implies that electrons scattered at low $Q^2$ and $x$ have to be
measured very close to the $e$ beam direction. Since there is no
upstream synchrotron radiation background the beam pipe radius can be
small ($~\simeq$ 2cm), and the pipe can be equipped with conical exit
windows.  In the THERA detector design sketched in Fig.\ref{tdet} the
mid-backward detector part can be removed for luminosity optimisation.
The highest luminosities require precooling in PETRA and dynamic
focusing. Much of the low $x$ physics, however, can be done with less
than 10~pb$^{-1}$ which opens the exciting possibility to perform the
low $x$ $ep$ programme simultaneously with precision electroweak
$e^+e^-$ physics.

THERA would complement the $e^+e^-$ and $pp$ colliders in the TeV
energy range and cover a huge field of physics as well as accessing
the smallest Bjorken $x$ values. A superconducting linac with a bunch
crossing time of a few 100~ns could also be combined with the Tevatron.
An electron ring of 60~GeV energy on top of the LHC would
access about the same kinematic region as THERA with an estimated
luminosity~\cite{keil} of $ \geq 10^{32}$cm$^{-2}$s$^{-1}$.

\section{Summary}
Deep inelastic lepton-nucleon scattering has been a most exciting
field of research over three decades. HERA has contributed enormously
to the development of strong interaction theory and it will continue
to do so with high precision data for quite some time. While many new
effects have been observed, they still give rise to basic, 
yet-unanswered
questions in the field of low $x$ physics. These regard the saturation of
the rise of $F_2$ and $xg$, the cross-section behaviour down to 
$x \sim 10^{-8}$, the mechanism of gluon emission, the nature of
diffraction, the theory of heavy quarks in the proton, the unknown
spin structure at low $x$, etc. Confinement is not understood, and
nearly all mass is carried by nucleons, the deep structure of which
has only just been addressed experimentally.  
Future experiments have to aim at maximum
accuracy, to extend the kinematic range, to reach regions of higher
quark and gluon density, to accelerate nuclei in colliding $eA$ mode
as well as to study polarised lepton-nucleon interactions at much
higher energy than hitherto. High-luminosity DIS has a discovery
potential which requires continuing efforts. These successfully focus
on three projects, HERA-3, the EIC and THERA, which all deserve to be
realised, since they cover different regions in phase space at
different times. One hundred years after Rutherford, 
lepton-nucleon
scattering experiments exploring the deep structure of matter will
continue to attract our attention.
%
%

\end{document}